\documentclass{article}

\usepackage{PRIMEarxiv}

\usepackage[utf8]{inputenc}
\usepackage[T1]{fontenc}
\usepackage{amsmath,amsfonts,amssymb}
\usepackage{graphicx}
\usepackage{booktabs}
\usepackage{microtype}
\usepackage{url}
\usepackage{hyperref}
\hypersetup{hidelinks}
\usepackage{xcolor}
\usepackage{tikz}
\usepackage{adjustbox}
\usepackage{placeins}
\usetikzlibrary{arrows.meta,positioning,shapes.geometric,shapes.misc}

\graphicspath{{./}}

\title{Distributed Air-Gap Flux and Rotor-Current Fusion for Operating-Regime Identification in a 10-MW Kaplan Hydrogenerator}

\author{%
Eduardo Jr Piedad$^{1,2,*}$, Rafel Roig$^{2}$, Xavier Escaler$^{2}$,\
Eduardo Prieto-Araujo$^{2}$, and Oriol Gomis-Bellmunt$^{2}$\\
$^{1}$DOST--Advanced Science and Technology Institute, Quezon City 1101 Philippines\\
$^{2}$Universitat Polit\`ecnica de Catalunya, 08034 Barcelona, Spain\\
$^{*}$Corresponding author: \texttt{eduardojr.piedad@asti.dost.gov.ph}
}

\begin{document}
\maketitle

\begin{abstract}
Reliable monitoring of hydroelectric generators requires descriptors that capture both electrical loading and electromagnetic field behavior. This work investigates operating-regime identification in the Porjus U9 10-MW Kaplan hydrogenerator using synchronized measurements from ten stator-mounted Hall probes and six rotor-current channels. Seven steady guide-vane-opening settings are considered, and each 300~s record is divided into 1~s windows. The resulting windows are represented by spatial Fourier descriptors of the circumferential air-gap field, probe-wise temporal flux indicators, and channel-wise RMS rotor-current features. Correlation analysis and principal component analysis are used to examine how the feature groups vary with the operating point, and Random Forest, radial-basis-function support vector classification, and multilayer perceptron models are evaluated for supervised identification of the guide-vane-opening state. The analysis shows that RMS rotor-current features mainly track the loading axis, while the magnetic-flux features reveal complementary information associated with spatial imbalance, waveform distortion, and weak low-frequency modulation. Spatial descriptors alone provide limited separability, yielding test accuracies below 27\%, whereas rotor-current features alone reach about 84--85\%. Combining flux and current information gives the most discriminative representation; the SVC-RBF model achieves 99.5\% test accuracy and macro-F1 score. The results indicate that distributed air-gap magnetic sensing, when fused with rotor-current measurements, can support accurate and interpretable data-driven monitoring of Kaplan hydrogenerator operating regimes.
\end{abstract}


\section{Introduction}
\label{sec:intro}

Hydropower is a key source of dispatchable renewable energy, and its role is increasingly important as power systems accommodate greater variability from renewable generation. This requires hydroelectric units to operate over wider hydraulic and electrical ranges, increasing the need for reliable condition monitoring under changing operating conditions. In Kaplan turbine-generator units, variations in guide vane opening (GVO) affect generator loading and may influence air-gap flux distortion, modulation, and spatial non-uniformity. These effects motivate signal-processing and machine-learning methods that can characterize operating states from measured generator signals.

Data-driven condition monitoring has been widely applied to hydropower and rotating machinery for anomaly detection, predictive maintenance, and operating-regime identification \cite{betti2021condition,lang2024hydropower,amini2025hybrid,lei2020applications}. Although deep learning has received significant attention, recent work has shown that simpler machine-learning methods can outperform deep models in motor fault detection when informative signal features are available \cite{briza2024simpler}. This supports the use of physically interpretable feature extraction combined with supervised classifiers, especially when transparency and compact representations are important.

For hydrogenerators, rotor-current measurements provide useful loading information but do not fully describe the spatial distribution of the air-gap magnetic field. Distributed magnetic sensing offers a more direct view of electromagnetic asymmetry, rotor eccentricity, field distortion, and air-gap non-uniformity. Previous studies have demonstrated the diagnostic relevance of air-gap flux measurements in hydrogenerators and synchronous generators \cite{babic2017fault,shaikh2022online}. However, the combined use of spatially distributed flux probes and rotor-current features for operating-state classification remains less explored.

This paper presents a spatiotemporal analysis and data-driven classification framework for a 10-MW Kaplan turbine generator using synchronized air-gap flux-density and rotor-current measurements. Ten stator-mounted Hall probes characterize the spatial and temporal behavior of the air-gap field, while six rotor-current channels provide loading-related information. From 1~s steady-state windows, spatial Fourier indicators, probe-resolved temporal flux features, and channel-wise RMS current features are extracted. Correlation analysis and PCA are used for interpretation, while supervised classifiers are used to evaluate GVO separability.

The objective of this work is to assess how spatial flux structure, temporal flux distortion, and rotor-current loading information contribute to the classification of Kaplan turbine generator operating conditions. The results show that current features primarily represent electrical loading, while flux-based features provide complementary spatial and modulation information. Their combination provides an effective representation for data-driven GVO classification and supports the use of distributed magnetic sensing for hydrogenerator condition monitoring.

\section{The 10-MW Generator}
\label{sec:generator}

This study uses measurements from the actual Porjus U9 Kaplan turbine generator. The generator configuration defines the spatial arrangement of the flux measurements, while the operating regimes provide labeled steady-state conditions for data-driven analysis.

\subsection{Generator Configuration}
\label{subsec:generator_configuration}

The Porjus U9 unit is a three-phase synchronous generator. Its main parameters are summarized in Table~\ref{tab:generator_parameters}. The machine has 10 poles and operates at a reference speed of 600~rpm, corresponding to a 50~Hz electrical frequency. The stator has 120 slots, an outer diameter of 3050~mm, an inner diameter of 2000~mm, and an axial length of 1450~mm. These parameters define the electromagnetic periodicity of the machine and provide the spatial reference for interpreting air-gap flux measurements around the stator.

Spatial magnetic measurements are particularly relevant for this generator because air-gap flux density is sensitive to non-uniform magnetic conditions around the machine circumference. Prior electromagnetic simulations under eccentric operation \cite{escaler_2024} indicate that voltage and flux-linkage waveforms are only weakly affected, whereas air-gap flux density, stator tooth forces, and torque show more evident asymmetry. This motivates the use of distributed flux measurements, together with rotor-current sensing, for monitoring operating-state behavior.

\begin{table}[!htbp]
    \centering
    \caption{Key parameters of the Porjus U9 synchronous generator.}
    \label{tab:generator_parameters}
    \begin{adjustbox}{max width=\linewidth}
    \begin{tabular}{lll}
        \hline
        \textbf{Parameter} & \textbf{Value} & \textbf{Unit} \\
        \hline
        Machine type & Three-phase synchronous & -- \\
        Number of poles & 10 & -- \\
        Reference speed & 600 & rpm \\
        Electrical frequency & 50 & Hz \\
        Stator outer diameter & 3050 & mm \\
        Stator inner diameter & 2000 & mm \\
        Stator axial length & 1450 & mm \\
        Number of stator slots & 120 & -- \\
        Stacking factor & 0.97 & -- \\
        Stator steel type & U9\_stator & -- \\
        \hline
    \end{tabular}
    \end{adjustbox}
\end{table}

\subsection{Flux and Current Measurements}
\label{subsec:dataset}

The dataset consists of synchronized air-gap flux-density and rotor-current signals acquired under steady-state operation. Flux density was measured using ten stator-mounted Hall-effect probes uniformly distributed around the stator, giving an angular spacing of \(36^\circ\); hence, the probe locations are defined as \(\theta_k = 36^\circ(k-1)\), for \(k=1,\dots,10\). Six rotor-current channels were recorded simultaneously to capture the corresponding excitation-current behavior. All channels were sampled at 250~Hz for seven guide vane opening (GVO) conditions, \(\{14^\circ,16^\circ,18^\circ,20^\circ,22^\circ,24^\circ,26^\circ\}\). Each operating condition contains a 300~s steady-state record. For subsequent analysis, each record is segmented into non-overlapping 1~s windows, corresponding to \(T=250\) samples per window. The flux signal from Hall probe \(k\) is denoted by \(B_k(t)\), and the rotor-current signal from channel \(j\) is denoted by \(i_j(t)\).

\section{Methodology}
\label{sec:methodology}

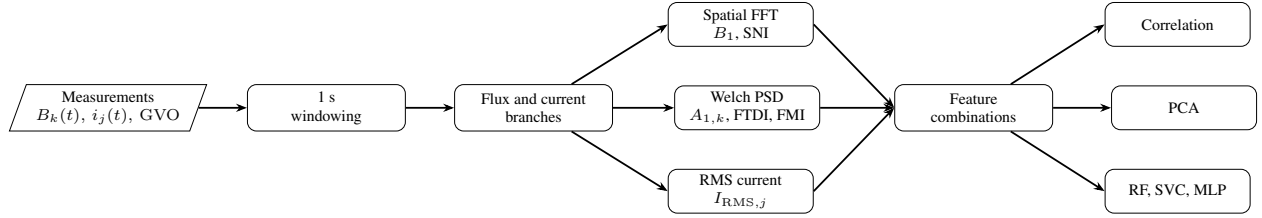
\begin{figure}[!htbp]
    \centering
    \scriptsize
    \resizebox{\linewidth}{!}{%
    \begin{tikzpicture}[
        node distance=0.62cm and 0.75cm,
        block/.style={rectangle, rounded corners, draw, align=center,
            minimum width=2.45cm, minimum height=0.72cm, inner sep=2.5pt},
        smallblock/.style={rectangle, rounded corners, draw, align=center,
            minimum width=2.25cm, minimum height=0.68cm, inner sep=2pt},
        line/.style={-{Stealth[length=1.7mm]}, thick},
        io/.style={trapezium, trapezium left angle=70, trapezium right angle=110,
            draw, align=center, minimum width=2.65cm, minimum height=0.72cm, inner sep=2.5pt}
    ]

        \node[io] (data) {Measurements\\$B_k(t),\, i_j(t),\, \mathrm{GVO}$};
        \node[block, right=of data] (window) {1~s\\windowing};
        \node[block, right=of window] (branches) {Flux and current\\branches};

        \node[smallblock, above right=0.58cm and 0.85cm of branches] (spatial)
        {Spatial FFT\\$B_1$, SNI};
        \node[smallblock, right=0.95cm of branches] (temporal)
        {Welch PSD\\$A_{1,k}$, FTDI, FMI};
        \node[smallblock, below right=0.58cm and 0.85cm of branches] (current)
        {RMS current\\$I_{\mathrm{RMS},j}$};

        \node[block, right=1.15cm of temporal] (features) {Feature\\combinations};

        \node[smallblock, above right=0.58cm and 0.80cm of features] (corr) {Correlation};
        \node[smallblock, right=0.90cm of features] (pca) {PCA};
        \node[smallblock, below right=0.58cm and 0.80cm of features] (ml) {RF, SVC, MLP};

        \draw[line] (data) -- (window);
        \draw[line] (window) -- (branches);
        \draw[line] (branches) -- (spatial.west);
        \draw[line] (branches) -- (temporal.west);
        \draw[line] (branches) -- (current.west);
        \draw[line] (spatial.east) -- (features.west);
        \draw[line] (temporal.east) -- (features.west);
        \draw[line] (current.east) -- (features.west);
        \draw[line] (features) -- (corr.west);
        \draw[line] (features) -- (pca.west);
        \draw[line] (features) -- (ml.west);

    \end{tikzpicture}%
    }
    \caption{The spatiotemporal analysis and classification framework.}
    \label{fig:methodology_flow}
\end{figure}

A signal-processing and learning framework shown in Fig.~\ref{fig:methodology_flow} is used to convert the synchronized flux-density and rotor-current measurements into features for GVO classification. Each 1~s window is processed through parallel flux and current branches. Flux signals are used to extract spatial and temporal indicators, while rotor-current signals are converted into channel-wise RMS features. The resulting features are analyzed using correlation analysis and PCA, and are then evaluated using supervised classifiers.

\FloatBarrier

\subsection{Window-Level Representation}
\label{subsec:window_representation}

Each 300~s steady-state record is divided into non-overlapping 1~s windows. Let \(w\) denote a window containing \(T=250\) samples. The windowed flux-density and rotor-current signals are denoted by
\begin{equation}
\begin{aligned}
    &B_k^{(w)}(t), \quad k=1,\dots,10,\\
    &i_j^{(w)}(t), \quad j=1,\dots,6,\quad t=1,\dots,T .
\end{aligned}
\end{equation}
Each window inherits the GVO label of the corresponding steady-state record and is treated as one sample for feature extraction and classification.

\subsection{Spatial Flux Features}
\label{subsec:spatial_flux_features}

Spatial flux information is extracted from the ten Hall probes distributed around the stator. For each window, the temporal mean of probe \(k\) is computed as
\begin{equation}
    \bar{B}_k^{(w)}
    =
    \frac{1}{T}
    \sum_{t=1}^{T} B_k^{(w)}(t),
    \quad k=1,\dots,10.
    \label{eq:window_flux_mean}
\end{equation}
The resulting vector \(\bar{\mathbf{B}}^{(w)} =
[\bar{B}_1^{(w)}, \bar{B}_2^{(w)}, \dots, \bar{B}_{10}^{(w)}]\)
represents one spatial flux profile around the stator. A discrete spatial Fourier transform is then applied:
\begin{equation}
    \hat{B}_n^{(w)}
    =
    \frac{1}{10}
    \sum_{k=1}^{10}
    \bar{B}_k^{(w)}
    e^{-j n \theta_k},
    \quad n=0,\dots,5.
    \label{eq:spatial_fft}
\end{equation}
The fundamental spatial amplitude and spatial non-uniformity index are defined as
\begin{equation}
\begin{aligned}
    B_1^{(w)} &= \left|\hat{B}_1^{(w)}\right|, \quad
    \mathrm{SNI}^{(w)}
    =
    \frac{
    \sqrt{\sum_{n=2}^{5}\left|\hat{B}_n^{(w)}\right|^2}
    }{
    B_1^{(w)}
    } .
\end{aligned}
\label{eq:spatial_features}
\end{equation}
Thus, \(B_1\) captures the dominant spatial component, while SNI quantifies higher-order spatial non-uniformity relative to the fundamental mode.

\subsection{Temporal Flux Features}
\label{subsec:temporal_flux_features}

Temporal flux dynamics are extracted independently from each Hall probe. For probe \(k\), Welch's method is used to estimate the power spectral density \(P_k^{(w)}(f)\). The dominant non-DC spectral component and its corresponding temporal fundamental amplitude are defined as
\begin{equation}
    f_{0,k}^{(w)}
    =
    \arg\max_{f>0} P_k^{(w)}(f),
    \qquad
    A_{1,k}^{(w)}
    =
    \sqrt{
    P_k^{(w)}
    \left(
    f_{0,k}^{(w)}
    \right)
    }.
    \label{eq:dominant_frequency_A1}
\end{equation}
The flux temporal distortion index is defined as
\begin{equation}
    \mathrm{FTDI}_k^{(w)}
    =
    \frac{
    \sqrt{
    \sum_f P_k^{(w)}(f)
    -
    P_k^{(w)}
    \left(
    f_{0,k}^{(w)}
    \right)
    }
    }{
    A_{1,k}^{(w)}
    },
    \label{eq:FTDI_probe}
\end{equation}
and the low-frequency modulation index is
\begin{equation}
    \mathrm{FMI}_k^{(w)}
    =
    \frac{
    \sum_{f \le 5~\mathrm{Hz}} P_k^{(w)}(f)
    }{
    \sum_f P_k^{(w)}(f)
    }.
    \label{eq:FMI_probe}
\end{equation}
Each window therefore yields 30 probe-resolved temporal flux features,
\[
    \{A_{1,k}^{(w)},\mathrm{FTDI}_k^{(w)},\mathrm{FMI}_k^{(w)}\}_{k=1}^{10}.
\]

\subsection{Rotor-Current Features}
\label{subsec:current_features}

Rotor-current features are extracted from the same 1~s windows. For each current channel, the RMS value is computed as
\begin{equation}
    I_{\mathrm{RMS},j}^{(w)}
    =
    \sqrt{
    \frac{1}{T}
    \sum_{t=1}^{T}
    \left(i_j^{(w)}(t)\right)^2
    },
    \quad j=1,\dots,6.
    \label{eq:Irms_channel}
\end{equation}
The current feature vector is
\begin{equation}
    \mathbf{I}_{\mathrm{RMS}}^{(w)}
    =
    [
    I_{\mathrm{RMS},1}^{(w)},\dots,
    I_{\mathrm{RMS},6}^{(w)}
    ].
    \label{eq:Irms_vector}
\end{equation}
Unlike a single averaged current descriptor, this representation preserves channel-wise current variation.

\subsection{Feature-Set Construction}
\label{subsec:feature_set_construction}

The extracted indicators are organized into spatial, temporal-flux, and current feature groups:
\begin{equation}
    \mathcal{X}_{\mathrm{S}}^{(w)}
    =
    \{
    B_1^{(w)},\mathrm{SNI}^{(w)}
    \},
    \label{eq:spatial_feature_set}
\end{equation}
\begin{equation}
\begin{aligned}
    \mathcal{X}_{\mathrm{T}}^{(w)}
    =
    \{&
    A_{1,1}^{(w)},\dots,A_{1,10}^{(w)},
    \mathrm{FTDI}_{1}^{(w)},\dots,\mathrm{FTDI}_{10}^{(w)}, \\
    &
    \mathrm{FMI}_{1}^{(w)},\dots,\mathrm{FMI}_{10}^{(w)}
    \},
\end{aligned}
\label{eq:temporal_feature_set}
\end{equation}
and
\begin{equation}
    \mathcal{X}_{\mathrm{I}}^{(w)}
    =
    \{
    I_{\mathrm{RMS},1}^{(w)},\dots,
    I_{\mathrm{RMS},6}^{(w)}
    \}.
    \label{eq:current_feature_set}
\end{equation}

Seven feature combinations are evaluated in the classification stage: spatial only \((\mathcal{X}_{\mathrm{S}})\), flux only \((\mathcal{X}_{\mathrm{S}}\cup\mathcal{X}_{\mathrm{T}})\), current only \((\mathcal{X}_{\mathrm{I}})\), current plus FTDI, current plus spatial \((\mathcal{X}_{\mathrm{I}}\cup\mathcal{X}_{\mathrm{S}})\), temporal flux plus current \((\mathcal{X}_{\mathrm{T}}\cup\mathcal{X}_{\mathrm{I}})\), and all features \((\mathcal{X}_{\mathrm{S}}\cup\mathcal{X}_{\mathrm{T}}\cup\mathcal{X}_{\mathrm{I}})\). Each feature vector is paired with its GVO label, giving a seven-class classification problem.

\subsection{Correlation Analysis}
\label{subsec:correlation_analysis}

Correlation analysis is used to quantify the monotonic relationship between each feature and GVO. Let \(x_m^{(w)}\) be the value of feature \(m\) in window \(w\), and let \(g^{(w)}\) be the corresponding numerical GVO label. The Pearson correlation coefficient is computed as
\begin{equation}
    \rho_m
    =
    \frac{
    \sum_{w=1}^{N}
    \left(
    x_m^{(w)}-\bar{x}_m
    \right)
    \left(
    g^{(w)}-\bar{g}
    \right)
    }{
    \sqrt{
    \sum_{w=1}^{N}
    \left(
    x_m^{(w)}-\bar{x}_m
    \right)^2
    }
    \sqrt{
    \sum_{w=1}^{N}
    \left(
    g^{(w)}-\bar{g}
    \right)^2
    }
    },
    \label{eq:pearson_corr}
\end{equation}
where \(N\) is the number of windows. Positive values indicate features that increase with GVO, negative values indicate features that decrease with GVO, and values near zero indicate weak monotonic association. This analysis is used for feature interpretation, not for classifier training.

\subsection{Dimensionality Reduction}
\label{subsec:dimensionality_reduction}

PCA is applied to standardized features to visualize the structure of the operating-state manifold. Given a standardized feature vector \(\mathbf{z}^{(w)}\), the \(i\)th principal-component score is
\begin{equation}
    \mathrm{PC}_i^{(w)}
    =
    \mathbf{w}_i^\top
    \mathbf{z}^{(w)},
    \quad i=1,2,
    \label{eq:pca_projection}
\end{equation}
where \(\mathbf{w}_i\) is the \(i\)th principal direction. PCA is used only for visualization and interpretation; classifier training is performed using the standardized feature vectors.

\subsection{Classification Models and Evaluation}
\label{subsec:classification_models}

Three classifiers similar to \cite{briza2024simpler} are evaluated - Random Forest (RF), support vector classification with a radial-basis-function kernel (SVC-RBF), and a multilayer perceptron (MLP). All models are implemented using pipelines with Z-score standardization. A stratified 70/30 train--test split is used for holdout evaluation, while five-fold stratified cross-validation is used to assess generalization stability. Performance is reported using test accuracy, macro-averaged F1 score, and row-normalized confusion matrices. The classifier configurations are summarized in Table~\ref{tab:classifiers}.

\begin{table}[!htbp]
    \centering
    \caption{Classifier models and hyperparameters.}
    \label{tab:classifiers}
    \footnotesize
    \setlength{\tabcolsep}{3pt}
    \renewcommand{\arraystretch}{1.12}
    \begin{tabular*}{\columnwidth}{@{\extracolsep{\fill}} l p{0.68\columnwidth}}
        \hline
        \textbf{Model} & \textbf{Hyperparameters} \\
        \hline
        Random Forest &
        300 trees; unlimited depth; bootstrap enabled \\
        SVC-RBF &
        \(C=10\); \(\gamma=\mathrm{scale}\); RBF kernel \\
        MLP &
        Hidden layers \((50,25)\); ReLU; Adam optimizer; max\_iter = 1000 \\
        \hline
    \end{tabular*}
\end{table}
\FloatBarrier

\section{Results and Discussion}
\label{sec:results_discussion}

This section presents the behavior of the extracted flux and current indicators, followed by correlation analysis, PCA visualization, and supervised GVO classification.

\subsection{Spatial Flux and Current Distributions}
\label{subsec:polar_flux_current}

\begin{figure}[!htbp]
    \centering
    \includegraphics[width=\linewidth]{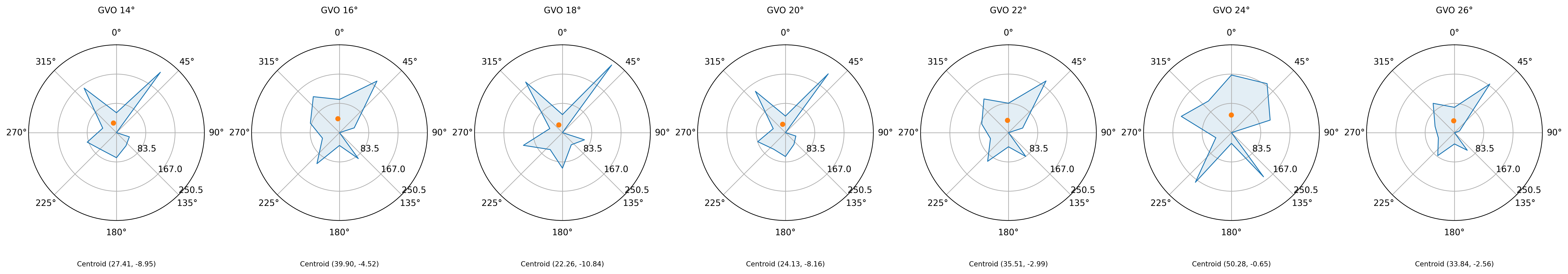}

    \vspace{0.3cm}

    \includegraphics[width=\linewidth]{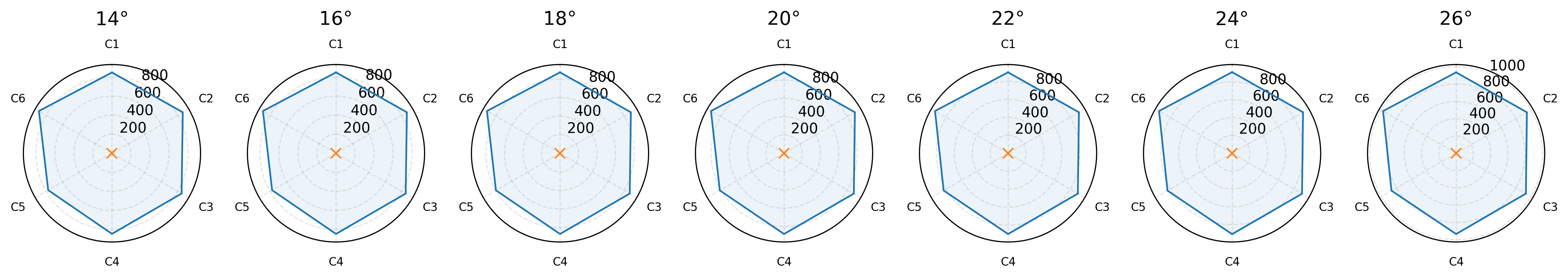}
    \caption{Polar representations of the mean air--gap flux distribution (in $mT$) at the upper row and the mean rotor-current distribution (in $A$) at the bottom row for all guide vane openings. The centroid indicates the effective center of each measured distribution.}
    \label{fig:polar_flux_current}
\end{figure}

Figure~\ref{fig:polar_flux_current} shows the polar representations of the mean air-gap flux distribution and the mean rotor-current distribution for all GVOs. The flux polygons are obtained from the ten Hall probes, while the current polygons are obtained from the six RMS current channels. The centroid of each polygon indicates the effective center of the measured distribution.

\FloatBarrier

The flux distributions preserve a similar angular pattern across GVOs, indicating that the dominant spatial magnetic structure is largely invariant with operating condition. The slight displacement of the flux-polygon centroid from the geometric origin suggests that the effective magnetic center does not perfectly coincide with the mechanical center. This may reflect a non-uniform air gap or fixed geometric asymmetry. Since the centroid displacement remains small and nearly unchanged across GVOs, the observed asymmetry is interpreted as structural rather than load-induced.

The current polar distributions show a clearer magnitude increase with GVO while preserving a similar channel-to-channel pattern. Thus, the current measurements mainly encode loading information. The probe-resolved temporal fundamental flux amplitudes exhibit similar spatial coherence and persistent probe-to-probe offsets, which are consistent with fixed spatial effects such as probe placement, slotting, winding layout, or air-gap non-uniformity.

\subsection{Indicator Trends Across Operating Conditions}
\label{subsec:indicator_trends}

\begin{figure}[!htbp]
    \centering

    \begin{minipage}[t]{0.48\linewidth}
        \centering
        \includegraphics[width=\linewidth]{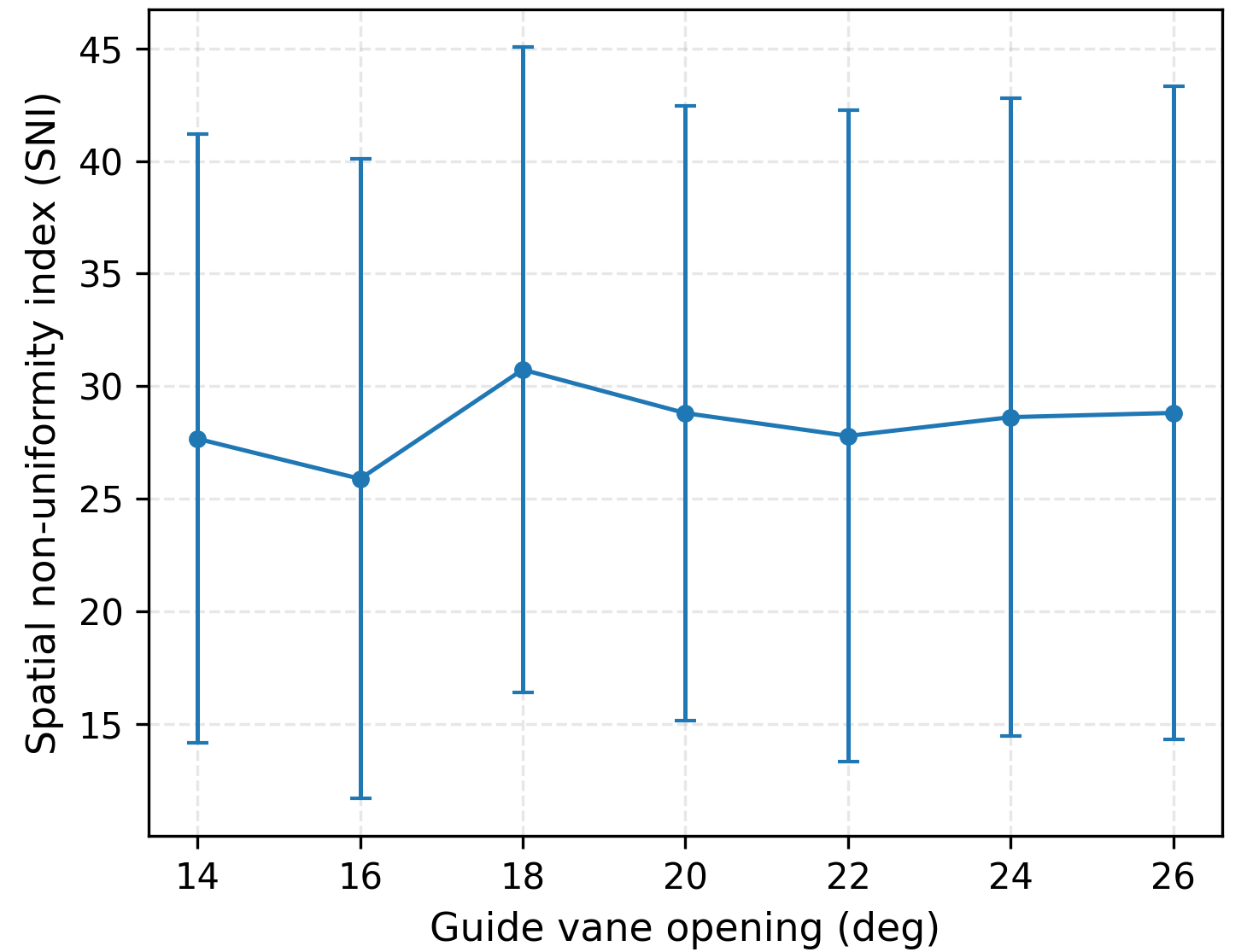}
        \vspace{-0.12cm}

        \small (a) SNI
    \end{minipage}
    \hfill
    \begin{minipage}[t]{0.48\linewidth}
        \centering
        \includegraphics[width=\linewidth]{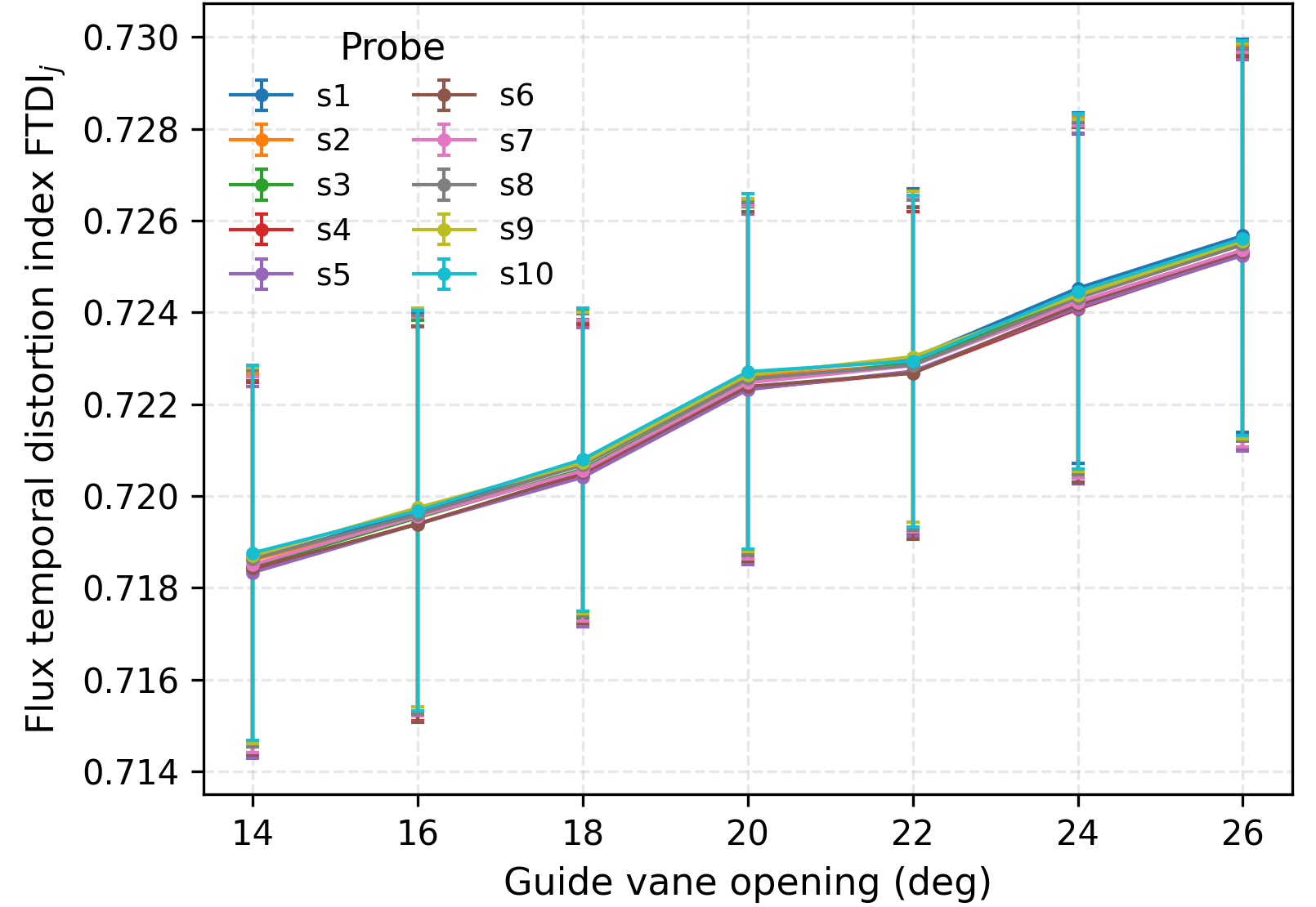}
        \vspace{-0.12cm}

        \small (b) FTDI
    \end{minipage}

    \vspace{0.30cm}

    \begin{minipage}[t]{0.48\linewidth}
        \centering
        \includegraphics[width=\linewidth]{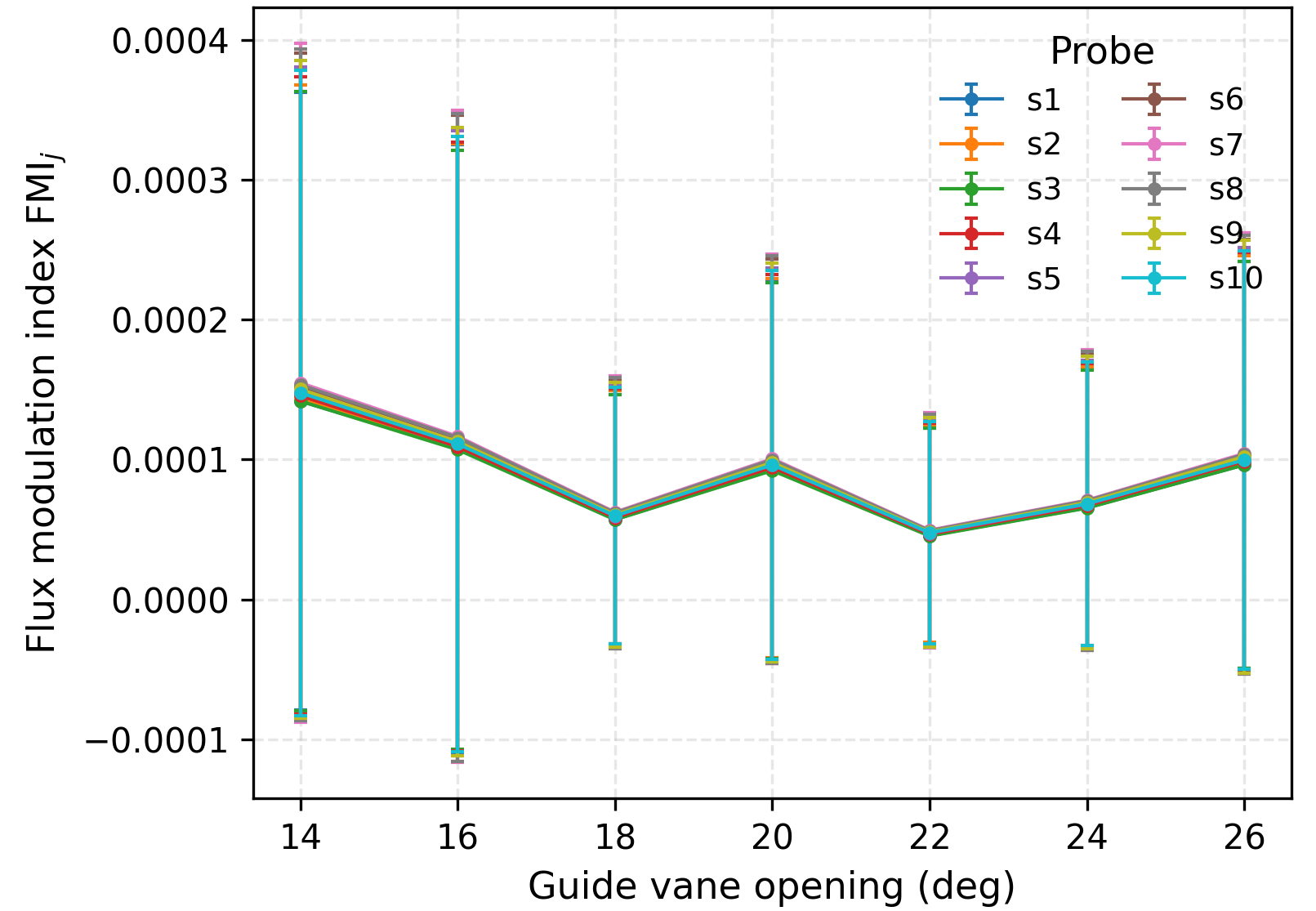}
        \vspace{-0.12cm}

        \small (c) FMI
    \end{minipage}
    \hfill
    \begin{minipage}[t]{0.48\linewidth}
        \centering
        \includegraphics[width=\linewidth]{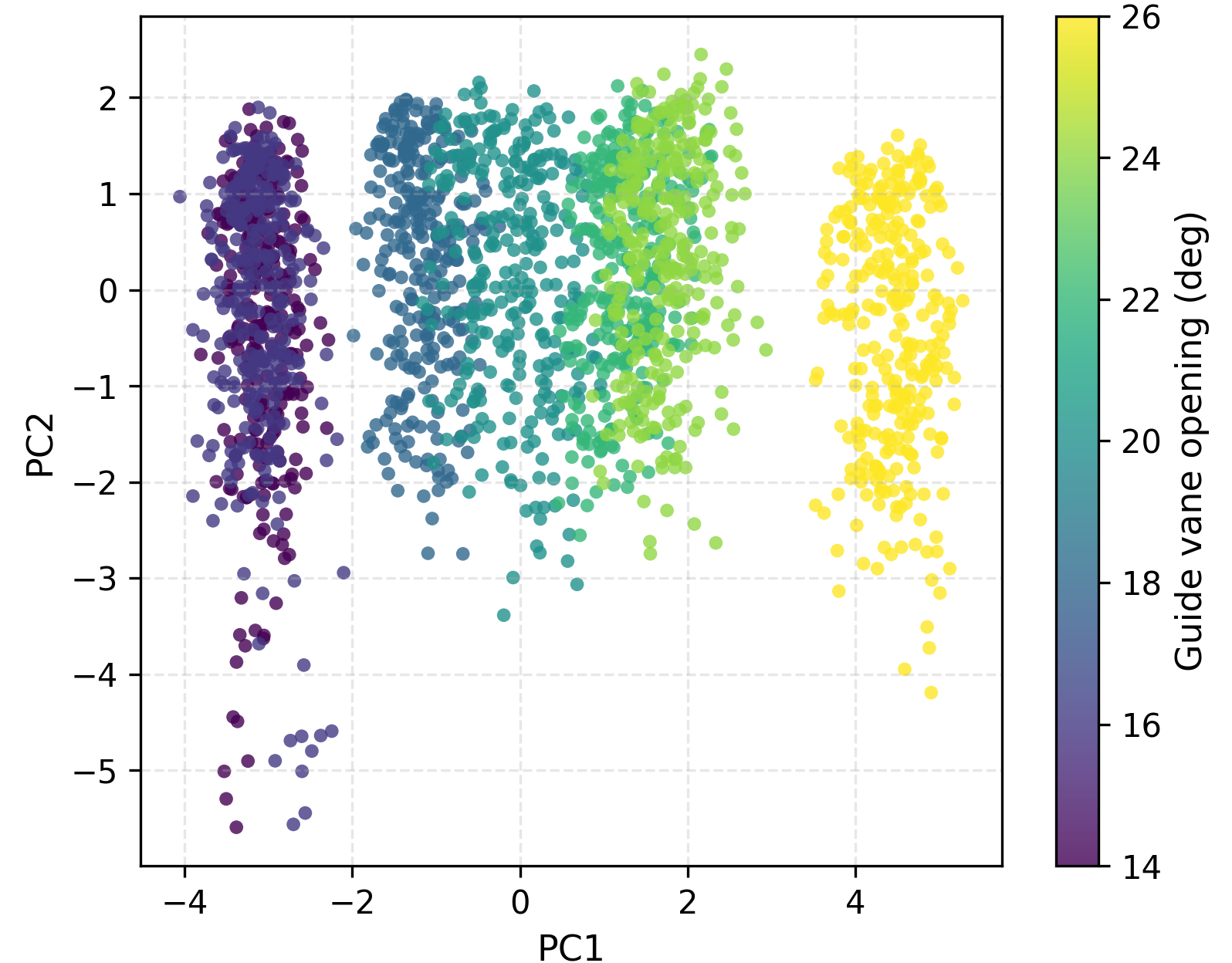}
        \vspace{-0.12cm}

        \small (d) PCA visualization
    \end{minipage}

    \caption{Summary of feature trends and low-dimensional structure: (a) spatial non-uniformity index, (b) probe-resolved flux temporal distortion index, (c) probe-resolved flux modulation index, and (d) PCA scores of the spatiotemporal flux and current feature set.}
    \label{fig:feature_trends_pca}
\end{figure}

Table~\ref{tab:indicator_summary} summarizes the main flux indicators per GVO, with temporal quantities averaged across the ten probes for compactness. The classifiers use the full probe-resolved feature vectors. The weak GVO dependence of \(B_1\) and SNI indicates that the spatial flux pattern is dominated by persistent machine characteristics rather than load-driven redistribution. In particular, Fig.~\ref{fig:feature_trends_pca}(a) shows that SNI remains large but nearly non-monotonic, suggesting structural spatial non-uniformity.

\begin{table}[!htbp]
    \centering
    \caption{Summary of flux and current indicators per GVO. Temporal flux and current quantities are averaged across probes/channels for compact reporting.}
    \label{tab:indicator_summary}
    \begin{adjustbox}{max width=\linewidth}
    \begin{tabular}{c c c c c c}
        \hline
        \textbf{GVO} & \(\boldsymbol{B_1}\) & \textbf{SNI} & \(\boldsymbol{\bar{A}_1}\) & \(\boldsymbol{\overline{\mathrm{FTDI}}}\) & \(\boldsymbol{\overline{\mathrm{FMI}}}\) \\
        \((^\circ)\) & (mT) & (--) & (PSD\(^{1/2}\)) & (--) & \((\times 10^{4})\) \\
        \hline
        14 & 18.58 & 27.66 & 841.24 & 0.7185 & 1.48 \\
        16 & 18.90 & 25.88 & 838.63 & 0.7196 & 1.12 \\
        18 & 17.67 & 30.73 & 840.21 & 0.7206 & 0.60 \\
        20 & 17.68 & 28.80 & 839.31 & 0.7225 & 0.96 \\
        22 & 17.67 & 27.79 & 840.03 & 0.7229 & 0.48 \\
        24 & 17.88 & 28.62 & 837.06 & 0.7243 & 0.68 \\
        26 & 17.16 & 28.81 & 847.36 & 0.7255 & 1.00 \\
        \hline
    \end{tabular}
    \end{adjustbox}
\end{table}

\begin{table}[!htbp]
    \centering
    \caption{Mean RMS current per GVO, averaged across six channels.}
    \label{tab:current_summary}
    \footnotesize
    \setlength{\tabcolsep}{3.5pt}
    \begin{adjustbox}{max width=\linewidth}
    \begin{tabular}{c c c c c c c c}
        \hline
        GVO \((^\circ)\) & 14 & 16 & 18 & 20 & 22 & 24 & 26 \\
        \hline
        \(\overline{I}_{\mathrm{RMS}}\) (A)
        & 844.5 & 844.9 & 867.3 & 880.7 & 897.4 & 904.5 & 932.1 \\
        \hline
    \end{tabular}
    \end{adjustbox}
\end{table}

By contrast, the current summary in Table~\ref{tab:current_summary} increases with GVO and therefore provides a direct loading-related descriptor. The temporal distortion index in Fig.~\ref{fig:feature_trends_pca}(b) also increases consistently with GVO across probes, indicating that the harmonic content of the air-gap flux becomes more pronounced as loading increases. The flux modulation index in Fig.~\ref{fig:feature_trends_pca}(c) remains small in magnitude, but its systematic variation across GVOs suggests sensitivity to weak low-frequency electromagnetic modulation.

\FloatBarrier

\subsection{Correlation Analysis}
\label{subsec:correlation_results}

Table~\ref{tab:correlation_summary} summarizes the Pearson correlation between each feature group and GVO. For probe- and channel-resolved indicators, the table reports the mean correlation across sensors or channels.

\begin{table}[!htbp]
    \centering
    \caption{Pearson correlation between feature groups and GVO.}
    \label{tab:correlation_summary}
    \scriptsize
    \setlength{\tabcolsep}{6.2pt}
    \renewcommand{\arraystretch}{1.15}
    \begin{adjustbox}{max width=\linewidth}
    \begin{tabular}{l c c p{4.15cm}}
        \hline
        \textbf{Feature} & \textbf{No.} & \textbf{Mean} & \textbf{Remark} \\
        \hline
        \(\boldsymbol{I_{\mathrm{RMS}}}\) & \textbf{6} & \textbf{0.978} & \textbf{Very strong monotonic relation} \\
        FTDI & 10 & 0.510 & Moderate monotonic relation \\
        \(B_1\) & 1 & -0.246 & Weak relation \\
        Flux \(A_1\) & 10 & 0.277 & Very weak / structural \\
        FMI & 10 & -0.109 & Very weak / structural \\
        \(\boldsymbol{\mathrm{SNI}}\) & \textbf{1} & \textbf{0.030} & \textbf{Very weak / structural} \\
        \hline
    \end{tabular}
    \end{adjustbox}
\end{table}

The correlation results confirm the different roles of current and flux-based indicators. The RMS current features have the strongest monotonic relationship with GVO, with a mean Pearson correlation of \(0.978\), confirming that rotor current is the primary electrical loading descriptor. FTDI shows a moderate positive correlation with GVO, indicating that temporal flux distortion captures load-dependent electromagnetic effects.

In contrast, \(B_1\), \(A_1\), FMI, and SNI exhibit weak correlations with GVO. The near-zero correlation of SNI is particularly important: although SNI is large in magnitude, it is not load-sensitive. This supports the interpretation that spatial non-uniformity mainly reflects persistent machine geometry, air-gap asymmetry, probe placement, or other structural effects rather than changes in GVO.

\subsection{PCA-Based Feature-Space Analysis}
\label{subsec:pca_results}

PCA was applied to the standardized fused feature set to examine the low-dimensional structure of the operating states. The first two principal components explain \(39.77\%\) and \(30.32\%\) of the total variance, respectively, giving a combined explained variance of \(70.09\%\). As shown in Fig.~\ref{fig:feature_trends_pca}(d), the GVOs form an ordered trajectory in the PCA plane, indicating that the extracted indicators encode a continuous operating manifold.

Table~\ref{tab:pca_loadings} summarizes the PCA loadings by feature group. For the probe-resolved temporal flux features, the table reports the mean loading across the ten Hall probes. For the RMS current features, it reports the mean loading across the six current channels.

\begin{table}[!htbp]
    \centering
    \caption{Feature representations in both PC loadings.}
    \label{tab:pca_loadings}
    \scriptsize
    \setlength{\tabcolsep}{7.2pt}
    \renewcommand{\arraystretch}{1.28}
    \begin{adjustbox}{max width=\linewidth}
    \begin{tabular}{l c c p{2.85cm}}
        \hline
        \textbf{Feature} & \(\boldsymbol{\mathrm{PC}_1}\) & \(\boldsymbol{\mathrm{PC}_2}\) & \textbf{Remark} \\
        \hline
        Mean \(I_{\mathrm{RMS}}\) & \textbf{0.389} & 0.008 & Electrical loading axis \\
        \(\overline{\mathrm{FTDI}}\) & 0.219 & -0.160 & Follows loading trend \\
        \(\bar{A}_1\) & 0.173 & -0.317 & Flux-related variance \\
        \(B_1\) & -0.116 & -0.361 & Spatial variation \\
        SNI & 0.018 & \textbf{0.676} & Spatial non-uniformity \\
        \(\overline{\mathrm{FMI}}\) & -0.030 & \textbf{-0.535} & Modulation structure \\
        \hline
    \end{tabular}
    \end{adjustbox}
\end{table}
\FloatBarrier

The loading structure indicates that \(\mathrm{PC}_1\) is primarily an electrical loading axis, strongly governed by the rotor-current RMS features. Although FTDI contributes positively to \(\mathrm{PC}_1\), its loading is smaller than that of the current features; therefore, it follows the loading trend but does not independently define the dominant principal direction.

In contrast, \(\mathrm{PC}_2\) is mainly a spatial/modulation axis. It is dominated by SNI and FMI, with additional contributions from \(B_1\) and \(\bar{A}_1\). These features structure the variance orthogonal to the main loading direction, showing that spatial non-uniformity and low-frequency modulation encode information not captured by rotor current alone.

\begin{table}[!htbp]
    \centering
    \caption{Classification performance of feature combinations.}
    \label{tab:classification_results}
    \footnotesize
    \setlength{\tabcolsep}{3pt}
    \begin{adjustbox}{max width=\linewidth}
    \begin{tabular}{l l c c c}
        \hline
        \textbf{Classifier} & \textbf{Feature set} & \textbf{CV accuracy} & \textbf{Test accuracy} & \textbf{Macro F1} \\
        \hline
        RF & Spatial only & 0.222 & 0.225 & 0.226 \\
        RF & Flux only & 0.768 & 0.778 & 0.777 \\
        RF & Current only & 0.845 & 0.854 & 0.854 \\
        RF & Current + FTDI & 0.850 & 0.857 & 0.856 \\
        RF & Current + spatial & 0.848 & 0.854 & 0.854 \\
        \textbf{RF} & \textbf{Flux + current} & \textbf{0.961} & \textbf{0.967} & \textbf{0.967} \\
        \textbf{RF} & \textbf{All features} & \textbf{0.976} & \textbf{0.983} & \textbf{0.983} \\
        \hline
        SVC-RBF & Spatial only & 0.246 & 0.237 & 0.236 \\
        SVC-RBF & Flux only & 0.824 & 0.795 & 0.792 \\
        SVC-RBF & Current only & 0.858 & 0.844 & 0.844 \\
        SVC-RBF & Current + FTDI & 0.853 & 0.843 & 0.843 \\
        SVC-RBF & Current + spatial & 0.847 & 0.835 & 0.835 \\
        \textbf{SVC-RBF} & \textbf{Flux + current} & \textbf{0.997} & \textbf{0.995} & \textbf{0.995} \\
        \textbf{SVC-RBF} & \textbf{All features} & \textbf{0.995} & \textbf{0.987} & \textbf{0.987} \\
        \hline
        MLP & Spatial only & 0.258 & 0.262 & 0.257 \\
        MLP & Flux only & 0.874 & 0.883 & 0.883 \\
        MLP & Current only & 0.850 & 0.838 & 0.838 \\
        MLP & Current + FTDI & 0.848 & 0.829 & 0.828 \\
        MLP & Current + spatial & 0.844 & 0.843 & 0.843 \\
        \textbf{MLP} & \textbf{Flux + current} & \textbf{0.995} & \textbf{0.989} & \textbf{0.989} \\
        \textbf{MLP} & \textbf{All features} & \textbf{0.994} & \textbf{0.987} & \textbf{0.987} \\
        \hline
    \end{tabular}
    \end{adjustbox}
\end{table}
\FloatBarrier

\subsection{Machine Learning Classification}
\label{subsec:classification_results}

The GVO classification performance was evaluated using seven feature combinations: spatial-only, flux-only, current-only, current plus FTDI, current plus spatial, flux plus current, and all features. Table~\ref{tab:classification_results} summarizes the five-fold cross-validation accuracy, holdout test accuracy, and macro-averaged F1 score for the three classifiers.

Spatial-only features are insufficient for GVO classification, with test accuracies below \(0.27\), indicating that \(B_1\) and SNI alone cannot resolve the seven operating states. Current-only features provide a stronger baseline of \(0.84\)--\(0.85\), but still confuse neighboring low-load conditions, particularly \(14^\circ\) and \(16^\circ\). Adding only FTDI or spatial features to current does not consistently improve performance, whereas flux-only features perform better than spatial-only features, showing the value of probe-resolved temporal flux information.

\begin{figure}[!htbp]
    \centering
    \includegraphics[width=\linewidth]{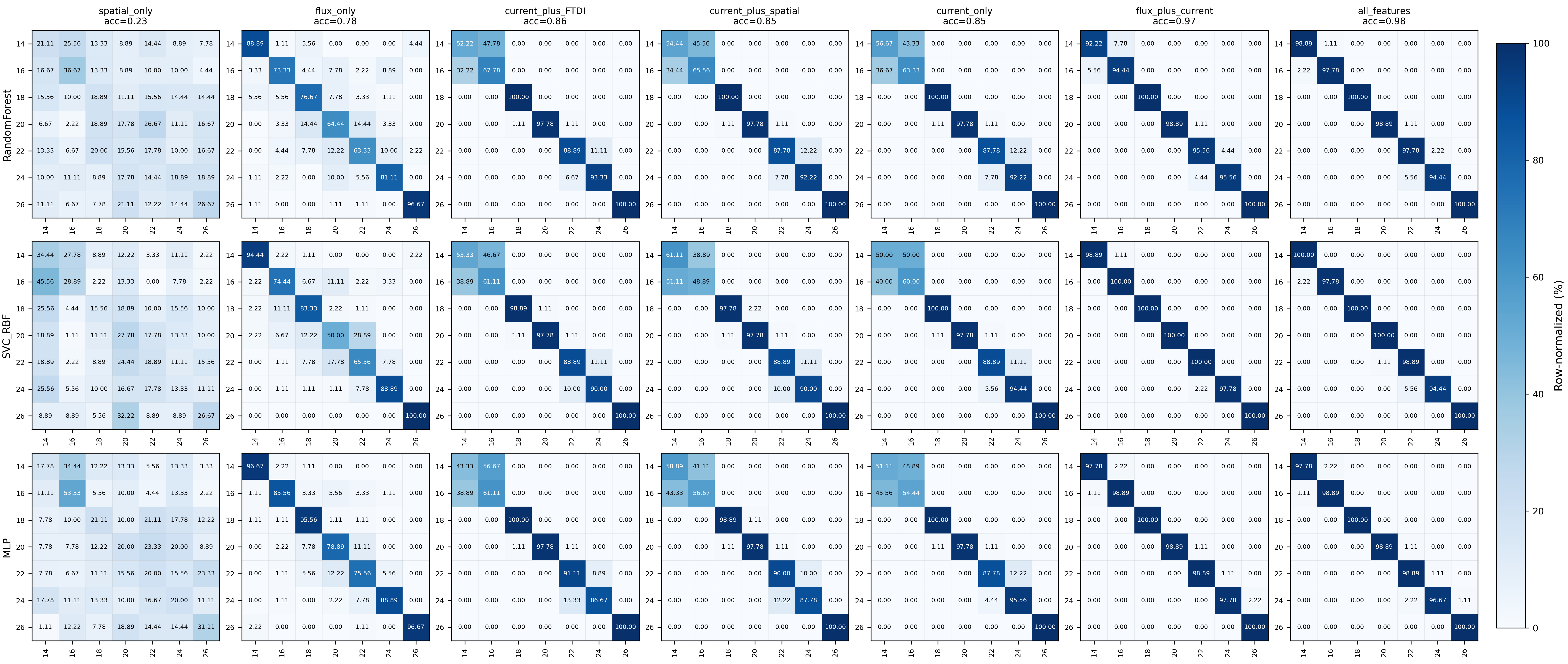}
    \caption{Combined row-normalized confusion matrices for the evaluated classifiers and feature combinations.}
    \label{fig:combined_confusion_matrix}
\end{figure}

The best results are obtained from combining flux-current features. SVC-RBF achieves the highest test accuracy and macro F1 score of \(0.995\), while the all-features setting does not consistently improve further, suggesting some redundancy. The confusion matrices in Fig.~\ref{fig:combined_confusion_matrix} confirm this trend: spatial-only predictions are diffuse, current-only models retain low-load confusion, and fused flux-current models produce the clearest diagonal structure. This demonstrates that current features encode the loading trend, while flux features provide complementary spatial and temporal electromagnetic information for reliable Kaplan turbine generator operating-state classification.

\FloatBarrier

\section{Conclusion}
\label{sec:conclusion}

This study presented a spatiotemporal signal-processing and data-driven classification framework for operating-condition monitoring of a 10-MW Kaplan turbine generator. Using synchronized air-gap flux-density and rotor-current measurements, the proposed approach extracted spatial flux indicators, probe-resolved temporal flux features, and channel-wise RMS current features from steady-state operating windows.

The analysis showed that the spatial flux distribution provides useful information about persistent machine asymmetry, while temporal flux indicators capture operating-dependent electromagnetic distortion and modulation. Rotor-current features primarily represent the electrical loading condition. Correlation analysis and PCA further confirmed that current and flux features describe complementary aspects of generator behavior: current features define the dominant loading direction, whereas flux-based features encode spatial and modulation structure.

The machine learning based classification results demonstrate that combining flux and current features provides the most reliable representation for distinguishing GVO operating states. This confirms the value of using distributed magnetic sensing together with rotor-current measurements for generator monitoring. The proposed framework offers a compact basis for data-driven operating-state identification and can support future condition-monitoring and digital-twin applications for hydrogenerator systems.

\bibliographystyle{IEEEbib}
\bibliography{strings,refs}

\end{document}